\documentstyle[floats,aps]{revtex}
\begin{document}
\def \beq{\begin{equation}}
\def \eeq{\end{equation}}
\def \beqarr{\begin{eqnarray}}
\def \eeqarr{\end{eqnarray}}
\def \be{\begin{equation}}
\def \ee{\end{equation}}
\def \bea{\begin{eqnarray}}
\def \eea{\end{eqnarray}}
\def \ta{{\tilde\alpha}}
\def \tg{{\tilde g}}     
\twocolumn[\hsize\textwidth\columnwidth\hsize\csname @twocolumnfalse\endcsname
\title{Classical versus Quantum Transport near Quantum Hall
Transitions}
\author{Efrat Shimshoni}
\address{Department of Mathematics-Physics, Oranim--Haifa University,
Tivon 36006, Israel.}
\date{\today}
\maketitle
\begin{abstract}
Transport data near quantum Hall transitions are
interpreted by identifying two distinct conduction regimes. The ``classical'' 
regime, dominated by nearest neighbor hopping between localized conducting
puddles, manifests an activated--like resistivity formula, and the 
quantized Hall insulator behavior. At very low temperatures $T$, 
or farther from the critical
point, a crossover occurs to a ``quantum" transport regime dominated
by variable range hopping. The latter is characterized by a different
$T$--dependence, yet the dependence on filling fraction is 
{\it coincidentally} hard to distinguish.
\end{abstract}
\vskip2pc]
\narrowtext
Magneto--transport measurements in the vicinity of quantum Hall (QH) 
transitions provided over the recent years an extensive variety of data 
\cite{phas-tran,chi,dhi,dual,hlldual,reslaw,qhi}, which stimulated a 
considerable confusion. The present paper suggests a way to settle the apparent
disagreement between different experimental results..

The traditional point of view asserts that transport properties near
the transitions between QH plateaux, and from a QH liquid to the insulator, 
should reflect the proximity to a second order quantum 
phase transition. Correspondingly, the d.c. resistivity tensor 
$\rho_{ij}$ at a given filling fraction $\nu$ and temperature $T$
should be described by a universal function $f(X)$ of a single parameter:
\beq
\rho_{ij}=\rho_{ijc}f\left({\Delta\nu\over T^{\kappa}}\right)\; ,
\label{rscale}
\eeq
where $\Delta\nu=\nu-\nu_c$ (the deviation from the critical filling $\nu_c$),
and $\rho_{ijc}$, $\kappa$ are universal. Here $\kappa$ is a
combination of critical exponents, $\kappa=1/zx$, where $z$ is the 
dynamical exponent and $x$ characterizes the divergance of the 
correlation length near criticality: $\xi\sim|\Delta\nu|^{-x}$.
Theoretical studies have predicted $x=7/3$ \cite{bodo}; experimentally,
a number of groups have obtained data in remarkable consistency with the 
scaling ansatz, at different (integer as well as fractional) QH transitions
\cite{yscl}. These results are further supported by data showing scaling at
finite frequency \cite{fscl} and current \cite{cscl}, which moreover confirm
the theoretical prediction $x=7/3$ and yield $z\approx 1$ (indicating 
the relevance of Coulomb interactions \cite{phas-tran} near the quantum
critical point).

The validity of the single parameter scaling form  Eq. (\ref{rscale}) has 
been, however, recently challenged \cite{reslaw}. Motivated by an earlier work
\cite{dual}, which identified a surprisingly robust (duality) symmetry 
relating the current--voltage curves $I(V)$ in the QH liquid phase 
($\Delta\nu>0$ 
\cite{fn1}) to $V(I)$ in the  insulator at  $-\Delta\nu$, the parameter 
$\Delta\nu$ was marked as a relevant scaling variable \cite{dualt}. The Ohmic 
resistivity plotted as a function of $\Delta\nu$ is indeed fitted (in a wide
range of parameters) by the formula
\beq
\rho_{xx}={h\over e^2}\exp\left[-{\Delta\nu\over \nu_0(T)}\right]\; ,
\label{rlaw}
\eeq    
However, counter to the expectation, $\nu_0(T)$ does not scale as $T^\kappa$,
but rather exhibits a linear dependence on $T$:
\beq
\nu_0(T)=\alpha T+\beta\; .
\label{nu0}
\eeq
The resistivity law  Eqs. (\ref{rlaw}), (\ref{nu0}) holds in various different
samples, as well as in different transitions, including plateau--to--plateau
transitions \cite{hlldual} (with an appropriate definition of the analog of
$\rho_{xx}$). The  parameters $\alpha$ and $\beta$ are sample--dependent,
and define a ``saturation temperature'' $T_s=\beta/\alpha$ which ranges 
between $~0.05K$ and $\sim 0.5K$. 
Moreover, even if one ignores this saturation,
attributing it to incomplete cooling of the carriers, a linear scaling of
$\nu_0(T)$ with $T$ is inconsistent with any sensible theory for the 
quantum critical behavior. 
It should be noted, that a partial set of curves obeying  Eqs. 
(\ref{rlaw}), (\ref{nu0}) (corresponding to a restricted range of parameters) 
can be collapsed on a ``traditional'' scaling curve \cite{shahar}, with an 
exponent $0<\kappa<1$. This observation raises a serious doubt concerning the
interpretation of the data in earlier experimental works: the distinction 
between a pure power law $T^\kappa$ with $\kappa\sim0.4-0.5$, and an 
alternative function which interpolated between $\sim T$ and a constant, 
may turn out to be rather difficult. Nevertheless, a recent experimental result
\cite{CZ} indicates a cross--over from one behavior to another, which will be 
discussed later in this paper in more detail.

Another set of experimental observations that appear to be inconclusive involve
the Hall resistance in the high magnetic field insulator. While part of the
data \cite{chi} support the theoretical prediction of a ``Hall insulator 
phase'' \cite{KLZ}, in which $\rho_{xx}$ diverges in the limit $T\rightarrow 0$
yet  $\rho_{xy}$ behaves as in a classical conductor (linearly dependent on 
the magnetic field $B$), other data \cite{dhi} exhibit a tendency to 
divergence of $\rho_{xy}$ as well. Yet a third class of experimental data 
\cite{dual,qhi,qhirus} have established the existance of a ``quantized Hall 
insulator'' (QHI) behavior in the insulator close to a fundamental QH state
($1/k$ with $k$ an odd integer): in this regime,  $\rho_{xy}$ is not only 
finite as asserted in \cite{KLZ}, but moreover maintains the quantized plateau
value $kh/e^2$. This phenomenon is a specific manifestation of the 
``semi--circle law'' \cite{ruzin}, which however extends beyond the range of 
validity expected from that theory. In particular, the observation of a QHI
behavior in the non--linear response regime \cite{dual} indicates a surprising
robustness of the phenomena, and has provided support to the idea that it is
intimately related to the validity of duality symmetry \cite{dualt}. 

In a previous work \cite{SA}, 
Shimshoni and Auerbach proposed a transport mechanism consistent
with the above described QHI phenomenon. The mechanism involves hopping 
across the junction between edge--states surrounding nearest neighbors 
in a random network of $1/k$-QH 
liquid puddles, carried out by quantum tunneling assisted by the temperature 
and current bias. Then, neglecting the quantum interference between different 
junctions in the network, it is proven that the Hall resistance is quantized
at the value dictated by the QH liquid, irrespective of the details of the 
longitudinal resistance associated with the hopping processes in the 
junctions. Note that this network model can be extended to the case where the
liquid puddles do not consist of a single type (a situation that is likely to
be applicable in an insulating regime close to more than one fundumental QH 
state), in which case $\rho_{xy}\sim B$ can be established (similarly to data
in Ref. \cite{chi}). It is later shown \cite{SAK}, that within this model one 
also obtains an activated--like behavior similar to  Eq. (\ref{rlaw}) (though
with $\nu_0(T)$ interpolating between $\sim T$ and a constant in a way 
different from the linear expression Eq. (\ref{nu0})). The linear dependence
on $\Delta\nu$ in this model is attributed to the relation between 
area--fraction of the liquid and the barrier heights.

Underlying the above described model for the transport, there is an essential
assumption of a finite {\it dephasing length} $L_\phi$, beyond which quantum
interfernce terms are suppressed. As long as the size of a QH liquid puddle,
around which electronic edge--states are extended, is larger than $L_\phi$ --
the {\it classical} random resistor network model is justified. In this sense,
the transport regime dominated by nearest--neighbor hopping processes is 
``classical''. This is even though the resistance associated with a single 
junction in the network (and given by a Landauer formula, where the two 
neighboring puddles are regarded as macroscopic reservoirs), is possibly 
dictated by quantum tunneling through the barrier. 

The classical model of Ref. \cite{SA} is not obviously a unique scenario  
which yields a quantized Hall resistance away from the strict QH phase. To 
test this, 
in a recent work \cite{PA} Pryadko and Auerbach have examined the effect of 
quantum interference in the network on  $\rho_{xy}$, and found a deviation from
the quantized value. In the case where $L_\phi$ is much larger than a puddle 
size, $\rho_{xy}$ vs. $B$ indicates an exponential divergence towards the 
insulating regime. A similar trend is indicated in other recent numerical data
as well \cite{SW}. This suggests that (counter to the naive intuition!)
a classical scenario is actually a 
{\it necessary condition} for supporting a QHI behavior; it also follows, that
the range of parameters in which it is observed should coincide with the range
of validity of  Eqs. (\ref{rlaw}), (\ref{nu0}). 
 
A cross--over from the classical to a quantum transport regime is expected, 
when the size of the QH puddles becomes smaller than $L_\phi$. Then, the
nearest neighbor localized puddle is not necessarily the optimal destination
of a hopping electron. Typically, randomly distributed electronic states that 
are close in energy are not close in real space. As a consequence, the
dominant transport mechanism is variable range hopping (VRH) \cite{vrh}. In 
this regime, a typical hop occurs between localized states separated by a
distance $R_h(T)$, which minimizes the exponential suppression of the 
hopping probability due to the difference in both energy and real space. 
Assuming a Coulomb gap in the density of states \cite{ES}, 
this hopping length is given by  
\beq
R_h(T)\sim \left(\xi e^2\over\epsilon k_B T\right)^{1/2}\; ;
\label{hopl}
\eeq 
here $\xi$ is the localization length, and $\epsilon$ the dielectric constant.
The resulting expression for the longitudinal resistance in the insulator is
\beq
\rho_{xx}\sim \rho_0\exp\left[\left({T_0\over T}\right)^{1/2}\right]\; ,
\quad T_0={e^2\over k_B\epsilon\xi}\; .
\label{rhop}
\eeq
(Similarly, in the QH phase Eq. (\ref{rhop}) holds for $1/\rho_{xx}$; to avoid
confusion, in the rest of the paper the expressions for $\rho_{xx}$ 
correspond to the insulator.)

To convert  Eq. (\ref{rhop}) into a dependence on the filling fraction,  
note that the localization length $\xi$ (which by definition describes a 
typical cluster over which an electronic state is extended), coincides with
the correlation length which tends to diverge near the transition \cite{PS}:
\beq
\xi=\xi_0\left({|\Delta\nu|\over \nu_c}\right)^{-x}
\label{xidiv}
\eeq 
(where $\xi_0$ is the value of $\xi$ deep in the localized phase). 
The critical behavior Eq. (\ref{xidiv}) is valid for $\Delta\nu\ll \nu_c$, 
so that $\xi\gg \xi_0$. On the other hand, the mechanism of VRH dominates the 
transport as long as $\xi$ is finite and {\it smaller} than $L_\phi$. Provided
the latter is a few orders of magnitude larger than $\xi_0$, there is a range 
of parameters where  Eq. (\ref{rhop}) holds {\it in coincidence} with 
Eq. (\ref{xidiv}) \cite{crhop}. As a result, one obtains
\beq
\rho_{xx}\sim \rho_0\exp\left[\left({C|\Delta\nu|^x\over T}\right)^{1/2}
\right]\; ,
\quad C\equiv{e^2\over k_B\epsilon\xi_0\nu_c^x}\; .
\label{rqlaw}
\eeq

In the regime where  Eq. (\ref{rqlaw}) is applicable, the experimental data
exhibit three prominent features: (a) the scaling form Eq. (\ref{rscale}) is
recovered with $\kappa=1/x\approx 0.43$ (given that indeed $x=7/3$), and 
$f(X)\sim e^{(CX^x)^{1/2}}$; (b) at a given $\Delta\nu$, 
\beq
\log\rho_{xx}(T)\sim T^{-1/2}\; ;
\label{rvsT}
\eeq
and (c) isotherms plotted as a function of $\nu$ are of the form
\beq
\log\rho_{xx}(\nu)\sim |\Delta\nu|^{1.15}\; .
\label{rvsnu}
\eeq
Comparing  Eq. (\ref{rvsnu}) with the empirical resistivity law (\ref{rlaw}),
we observe that {\it by mere coincidence} (which stems from the specific 
value of the exponent $x$), the two functional forms are practically 
indistinguishable. Similarly, it is hard to distinguish the temperature
dependence  Eq. (\ref{rvsT}) from the fit to $T^{-\kappa}$ employed in Ref. 
\cite{CZ}; it is suggested that the VRH scenario is, in fact, a more 
appropriate basis for interpretation of the data in the low $T$ regime. 

As mentioned above, the VRH scenario is consistent in the regime where the
transport is quantum coherent, namely for $\xi<L_\phi$. Hence,  the quantum 
regime terminate once $\xi$ approaches $L_\phi$ due to either increase of 
temperature, or the divergence of $\xi$ sufficiently close
to $\nu_c$. To estimate the boundary of the corresponding region in parameter 
space, an explicite expression for $L_\phi$ is needed. It turns out 
\cite{vrh} that in the VRH regime, the length scale which plays the role of a 
dephasing length is the hopping length $R_h(T)$ (Eq. (\ref{hopl})). This 
implies that a cross--over to a ``classical'' transport regim occurs at 
$T\sim T_0$ (where $T_0$ is defined in  Eq. (\ref{rhop})). 
Note that this criterion is consistent with the observation, that for 
$T>T_0$ the longitudinal resistivity no longer indicates the exponential 
divergence characteristic of strong localization. Employing  Eq. 
(\ref{xidiv}) we conclude that for a fixed $\nu$, a cross--over to the 
classical regime occurs at a temperature $T_x$, where
\beq
T_x\sim T_0=
{e^2\over k_B\epsilon\xi_0}\left({|\Delta\nu|\over \nu_c}\right)^x\; .   
\label{Tx}
\eeq
Alternatively, for a fixed $T$, the cross--over occurs at 
$|\Delta\nu|_x$, where
\beq
{|\Delta\nu|_x\over \nu_c}
\sim \left({\epsilon\xi_0 k_BT\over e^2}\right)^{1/x}\; .   
\label{dnux}
\eeq
As argued in Ref. \cite{PS}, the latter expression defines the width of the 
peaks in $\sigma_{xx}$ near QH transitions. However, it should be emphasized 
that (at a fixed $T$) critical scaling of the data is expected to hold
{\it out side} this width, while $|\Delta\nu|<|\Delta\nu|_x$ corresponds to a
classical transport regime.

I next show that the data which clearly manifest the activated--like behavior
resistivity law (\ref{rlaw}), (\ref{nu0}) (Ref. \cite{reslaw}) and a QHI
behavior (Refs. \cite{qhi,qhirus}), mostly correspond to the classical regime
by the criterion suggested above. A quantitative estimate of $|\Delta\nu|_x$
from Eq. (\ref{dnux}) is possible provided the ``bare'' localization length 
$\xi_0$ is known. Unfortunately, this parameter can not be extracted 
independently from the available information about the samples. However, the
fact that the integer QH effect is observed indicates that the 
single--electron states are localized over a length scale at least as large as
the magnetic length $l=(\hbar c/eB)^{1/2}$. Hence, the insertion $\xi_0\sim l$
provides a {\it minimal} estimate of $|\Delta\nu|_x$ for a given $T$. In
Ref. \cite{reslaw}, close to the critical field in the InGaAs/InP sample
($B_c=2.14\, T$, corresponding to $\nu_c=0.562$ and carrier density 
$n=3\times 10^{10}\, cm^{-2}$), one gets $l\approx 170\AA$. The implied 
lower bound on the width of the classical regime is 
$(\Delta\nu)_x/\nu_c\sim \pm 0.2$ for the highest temperature isotherm 
($T=2.21\, K$), and $(\Delta\nu)_x/\nu_c\sim\pm 0.1$ for $T=0.3\, K$. Comparing
with the data, it turns out that the range of $\nu$'s where $\log\rho_{xx}$ 
vs. $\nu$ is strictly linear is not much larger than this lower bound. A more
conclusive statement can be made regarding the quantized
Hall resistance data of Hilke {\it et al.} in Ref \cite{qhi}: 
there, $l\approx  130\AA$ which implies that
at the lowest displayed temperature ($T\approx 0.3\, K$), the classical regime
extends at least within $\Delta B/B_c\sim 0.09$. Indeed, this estimate implies
an upper field $B_u=B_c\times 1.09$ which roughly coincides with the field at
which the $\rho_{xy}$ data terminate (due to insufficient accuracy of the
measurement). Note that the range of observed QHI increases with $T$ or with 
an increased current bias, as long as a plateau in the QH phase is preserved.
Beyond a certain $T$, the quantization in the insulator is destroyed at the 
same time with the entire plateau, due to excitations to higher Landau levels.

To further test the central arguments of this paper, one should examine 
experimental data that extend over a wide enough range of temperatures below
and above the cross-over point $T_x$. The classical and quantum regimes are
then clearly distinct in terms of the $T$-dependence of  $\rho_{xx}$ for a
given $\nu$. The functional dependence on $\nu$ is, however, nearly identical:
$\log\rho_{xx}$ is expected to be approximately linear in $\nu$ in a wide range
of parameters extending over both regimes. This is possibly
a major source of confusion in the literature. In particular, in Ref. 
\cite{CZ} a cross--over in the $T$-dependence is clearly observed, however
the single cross--over point identified there ($T_x\approx 0.1\, K$) is an
average over a range of filling fractions. Similarly, it is possible that in
Ref. \cite{reslaw} as well, the outskirts of the range of $\Delta\nu$ 
indicating $\log\rho_{xx}\sim\nu$ extend into the quantum regime. The formula
Eq. (\ref{nu0}) is then not necessarily the only possible fit of the slope.

To summarize, I propose an interpretation of the extensive set of data close 
to QH transitions which distinguishes two conduction regimes. The classical
regime is established closer to the critical point and at relatively high $T$.
It is dominated by hopping between nearest neighbor hopping between conducting
QH puddles, whose typical size is larger than the dephasing length $L_\phi$.
Hence, the transport coefficients do not depend on $L_\phi$, but rather on
the details of the narrow junctions separating the puddles. When mapped to
a QH liquid--to--insulator transition, the characteristic behavior of the
resistivity tensor in this regime is an activated--like $\rho_{xx}$, and
quantization of $\rho_{xy}$ in the insulator. The quantum regime is established
at lower $T$ and farther from the critical point, where the transport is 
dominated by VRH. The limit of validity of VRH (corresponding to 
$\xi\sim R_h(T)$, where $R_h(T)$ is identified with $L_\phi$), provides an 
estimate of the boundary between the regimes (Eqs. (\ref{Tx}), (\ref{dnux})).
The classical and quantum regimes are very hard to distinguish by 
$\nu$--dependence of $\log\rho_{xx}$ ($\sim\Delta\nu$ vs. 
$\sim(\Delta\nu)^{1.15}$, respectively). It is predicted that the cross--over 
between them should be more clearly indicated by a change in the 
$T$--dependence, accompanied by a deviation of $\rho_{xy}$ in the insulator 
from a quantized plateau.  

\acknowledgements

I thank A. Auerbach, S. Girvin, M. Hilke, D. Huse, S. Murphy, D. Shahar, 
U. Sivan and S. Sondhi for useful conversations, and P. Coleridge for
informing me of his data prior to publication. This work was supported 
by grant no. 96--00294 from the United States--Israel Binational Science 
Foundation (BSF), Jerusalem, Israel.


\begin{references}
\bibitem{fn1}
In the case where the QH liquid state is {\it fractional}, $\nu$ is replaced by
the filling fraction of composite fermions; see Ref. \cite{dualt}.
\end{references}
\end{document}